\documentclass[fleqn,usenatbib]{mnras}

\usepackage{newtxtext,newtxmath}

\usepackage[T1]{fontenc}
\usepackage{ae,aecompl}

\usepackage{graphicx}	
\usepackage{amsmath}	
\usepackage{amssymb}	
\usepackage{subfig}

\newcommand{\cxo}{{\it Chandra}}
\def\xmm{\textit{XMM-Newton}}

\newcommand{\ergs}{erg\,s$^{-1}$}

\newcommand{\src}{MCSNR\,J0513-6724}
\newcommand{\osrc}{BSDL\,923}

\newcommand{\SII}{[S\,{\sc ii}]}

\usepackage{xcolor}

\title[ \src]{ Discovery of a very young high-mass X-ray binary associated with the supernova remnant \src\, in the LMC }

\author[C. Maitra et al.]{C. Maitra,$^{1}$\thanks{E-mail: cmaitra@mpe.mpg.de}
F. Haberl,$^{1}$
M. D. Filipovi\'c,$^{2}$
A. Udalski,$^{3,10}$
P.~J.~Kavanagh,$^{4}$
S. Carpano,$^{1}$
\newauthor
P.~Maggi,$^{5}$
M. Sasaki,$^{6}$
R.~P.~Norris$^{2,7}$
A. O'Brien$^{2,7}$
A.~Hotan$^{7}$
E.~Lenc$^{7}$
\newauthor
M.K. Szyma{\'n}ski$^{3,10}$
I. Soszy{\'n}ski$^{3,10}$
R. Poleski$^{3,8,10}$
K. Ulaczyk$^{3,9,10}$
\newauthor
P. Pietrukowicz$^{3,10}$
S. Koz{\l}owski$^{3,10}$
J. Skowron$^{3,10}$
P. Mr{\'o}z$^{3,10}$
K. Rybicki$^{3,10}$
\newauthor
P. Iwanek$^{3,10}$
M. Wrona$^{3,10}$
\\
$^{1}$Max-Planck-Institut f{\"u}r extraterrestrische Physik, Gie{\ss}enbachstra{\ss}e, 85748 Garching, Germany\\
$^{2}$Western Sydney University, Locked Bag 1797, Penrith South DC, NSW 1797, Australia\\
$^{3}$Astronomical Observatory, University of Warsaw, Aleje Ujazdowskie 4, 00-478 Warsaw, Poland\\
$^{4}$School of Cosmic Physics, Dublin Institute for Advanced Studies, 31 Fitzwillam Place, Dublin 2, Ireland\\
$^{5}$ Universit\'e de Strasbourg, CNRS, Observatoire astronomique de Strasbourg, UMR 7550, F-67000 Strasbourg, France\\
$^{6}$ Remeis Observatory and ECAP, Universit{\"a}t Erlangen-N{\"u}rnberg, Sternwartstr. 7, 96049 Bamberg, Germany\\
$^{7}$ CSIRO Astronomy and Space Science, PO Box 76, Epping, NSW 1710, Australia\\
$^{8}$ Department of Astronomy, Ohio State University, 140 W. 18th Ave.,
Columbus, OH~43210,~USA\\
$^9$ Department of Physics, University of Warwick, Gibbet Hill Road,
Coventry, CV4~7AL,~UK\\
$^{10}$ OGLE collaboration
}

\date{Accepted XXX. Received YYY; in original form ZZZ}

\pubyear{2019}
\begin{document}
\label{firstpage}
\pagerange{\pageref{firstpage}--\pageref{lastpage}}
\maketitle

\begin{abstract}
We report the discovery of a very young high-mass X-ray binary (HMXB) system associated with the supernova remnant (SNR) \src\ in the  Large Magellanic Cloud (LMC), using \xmm\ X-ray observations. The HMXB is located at the geometrical centre of extended soft X-ray emission, which we confirm as an SNR. The HMXB spectrum is consistent with an absorbed power law with spectral index $\sim$1.6 and a luminosity of $7\times10^{33}$ \ergs\ (0.2--12\,keV). Tentative X-ray pulsations are observed with a periodicity of 4.4\,s and the OGLE I-band light curve of the optical counterpart from more than 17.5\,years reveals a period of 2.2324$\pm$0.0003\,d, which we interpret as the orbital period of the binary system. 
The X-ray spectrum of the SNR is consistent with non-equilibrium shock models as expected for young/less evolved SNRs. From the derived ionisation time scale we estimate the age of the SNR to be $<$6\,kyr. The association of the HMXB with the SNR makes it the youngest HMXB, in the earliest evolutionary stage known to date. A HMXB as young as this can switch on as an accreting pulsar only when the spin period has reached a critical value. Under this assumption, we obtain an upper limit to the magnetic field of $<$5$\times10^{11}$\,G. This implies several interesting possibilities including magnetic field burial, possibly
by an episode of post-supernova hyper-critical accretion. Since these fields are expected to diffuse out on a timescale of $10^{3}-10^{4}$ years, the discovery of a very young HMXB can provide us the unique opportunity to observe the evolution of the observable magnetic field for the first time in X-ray binaries.
\end{abstract}




\section{Introduction}
Young neutron stars (NSs) residing in X-ray binary (XRB) systems are extremely rare objects and can provide unique insights on the birth properties and early evolution of NSs. In addition, these objects are ideal probes for the physics of accretion onto an NS at early evolutionary stages, and its implications on NS spins, magnetic fields, etc. Neutron stars are expected to be born rapidly spinning with periods around a few times 10 ms.  The spin period is expected to slow down over time by magnetic dipole braking and if the object is part of a binary system, by propeller effects at the onset of the accretion of matter from the companion \citep{1991PhR...203....1B}.  
 
Finding an NS XRB in its parent supernova remnant (SNR) implies a very young binary system, as the visibility time of SNR is only a few $10^4$~yr, much less than the lifetimes of high-mass XRBs. Associations of NS XRBs with SNRs were first proposed by \cite{1994AJ....107.1363H} from the ROSAT HRI data for two SNRs in the Small Magellanic Cloud. The first one, SNR\,IKT\,21 \citep{2004A&A...421.1031V} for which pulsations at 345\,s were  discovered by \cite{2000ApJ...531L.131I}, and a Be XRB candidate in the direction of IKT\,25 that could not be confirmed by either \xmm\ or \cxo.
For both of the above sources however, the physical association of the HMXB with the SNR is not clearly established as they are not located near the centre of the remnant.
Data from the \xmm\ and \cxo\  observatories have been instrumental in the discovery of only a few more such systems, and therefore finding an NS XRB with an SNR association continues to remain a rare discovery. This is mostly attributed to the fact that the timescale for which the remnant is visible is typically three orders of magnitude smaller than typical binary evolution timescale. Some secure XRB-SNR associations known are SXP\,1062 \citep{2012A&A...537L...1H,2018MNRAS.475.2809G}, DEM\,L241 \citep{2012ApJ...759..123S}, SXP\,1323 
\citep{2019MNRAS.485L...6G} in the Magellanic Clouds, and SS\,433 and Circinus\,X-1 in our Galaxy \citep{1980A&A....84..237G,2013ApJ...779..171H}. The youngest among them until now is Circinus\,X-1, with an estimated age $<$4600\,years \citep{2013ApJ...779..171H}. Finding the majority of these systems in the Magellanic Clouds (MCs) is not surprising given the ideal environment they have for hosting young stellar remnants, a high formation efficiency for high-mass X-ray binaries (HMXBs), as well as relatively small distance and low foreground absorption conducive for performing detailed studies. 
 
 We report here the discovery of a new HMXB at the geometrical centre of \src\ , which is confirmed as an SNR for the first time in this work.
 \src\ was previously a candidate SNR in the LMC \citep{Bozzetto2017}. It was first reported by \cite{1999A&AS..139..277H} recording an extent of 17.5\arcsec\ albeit a `hard' hardness-ratio. \cite{Bozzetto2017} found the presence of a weak \SII\ ring from Magellanic Cloud Emission Line Survey (MCELS) data and also the evidence for an unresolved radio source. The observations and their analysis are described in Sect. 2. Section 3 presents the results and Sect. 4 the discussions and conclusions.

\section{Observations and analysis}
\label{Sect:data}

\begin{table*} 
\caption{\xmm\ observations details of \src\ .} 
\begin{tabular}{ccccc} 
\hline
Date       & ObsID      & Exposure           & Off-axis angle         & Telecope vignetting\\ 
           &            & PN / MOS2 / MOS1   & PN / MOS2  / MOS1      & PN / MOS2 / MOS1 \\
yyyy/mm/dd &            & (ks)               &                        &                   \\ 
\hline 
2012/02/04 & 0671090101 & 23.9 / 30.0 / 30.2 &  4.0\arcmin / 3.1\arcmin / 3.6\arcmin & 0.86 / 0.88 /  0.86 \\ 
\hline 
\label{tabxray} 
\end{tabular} 
\end{table*} 

\subsection{X-ray observations and analysis}
\label{xray}
\src\ was observed serendipitously with \xmm\  in 2012. The observation details are given in Table~\ref{tabxray}. EPIC \citep{2001A&A...365L..18S,2001A&A...365L..27T} observations were processed with the \xmm, data analysis software SAS version 17.0.0\footnote{Science Analysis Software (SAS): http://xmm.esac.esa.int/sas/}. We searched for periods of high background flaring activity by extracting light curves in the energy range of 7.0--15.0\,keV and removed the time intervals with background rates $\geq$~8 and 2.5\,cts\,ks$^{-1}$~arcmin$^{-2}$ for EPIC-PN and EPIC-MOS respectively \citep{2013A&A...558A...3S}. Events were extracted using the SAS task \texttt{evselect} by applying the standard filtering criteria (\texttt{\#XMMEA\_EP \&\& PATTERN<=4} for EPIC-PN and \texttt{\#XMMEA\_EM \&\& PATTERN<=12} for EPIC-MOS).


\begin{figure*}
 \includegraphics[scale=0.28]{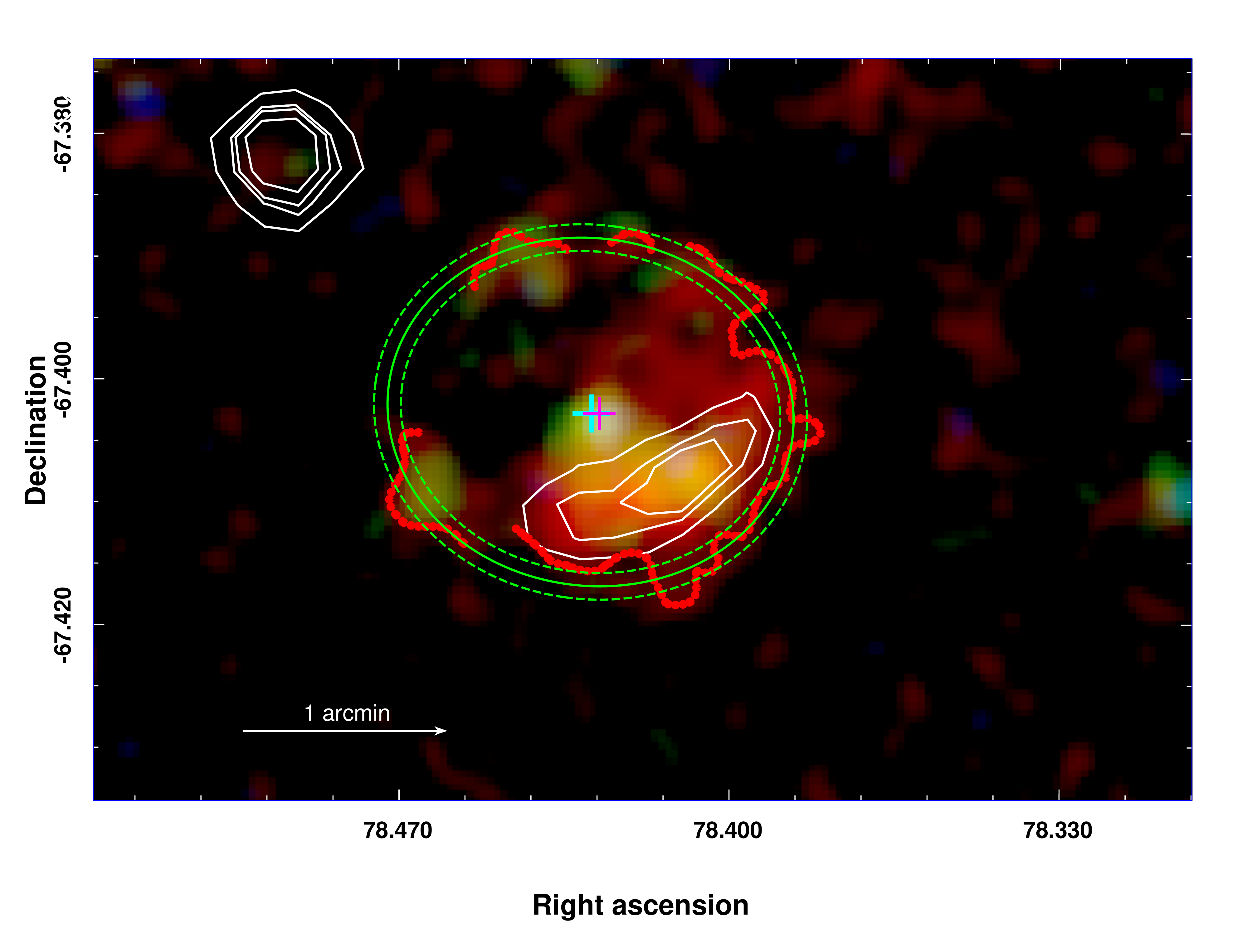}
   \caption{\xmm\ EPIC RGB (R=~0.2--1\,keV, G=~1--2\,keV, B=~2--4.5\,keV) image of \src. Overlaid are white radio contours from the latest Australian Square Kilometre Array Pathfinder (ASKAP) survey of the LMC at 888~MHz (bandwidth is 288~MHz). The radio continuum contours correspond to 1, 2, and 3~mJy~beam$^{-1}$ while the image beam size is 13.7\arcsec$\times$11.8\arcsec\ and local rms is $\sim$0.2~mJy~beam$^{-1}$. The full details of the ASKAP LMC survey including the data reduction and analysis will be presented in an upcoming paper by Filipovi{\'c} et al. The red dotted points indicate the contour level corresponding to 3$\sigma$ above the average background surface brightness. The green solid line shows the best-fit ellipse to the contour, with the dashed lines denoting the 1$\sigma$ errors on the best-fit. The cyan cross indicates the best-fit centre of the ellipse, and the optical position of the HMXB is shown in magenta.} 
   \label{fig1}
\end{figure*}

\begin{figure*}
    \includegraphics[scale=0.27]{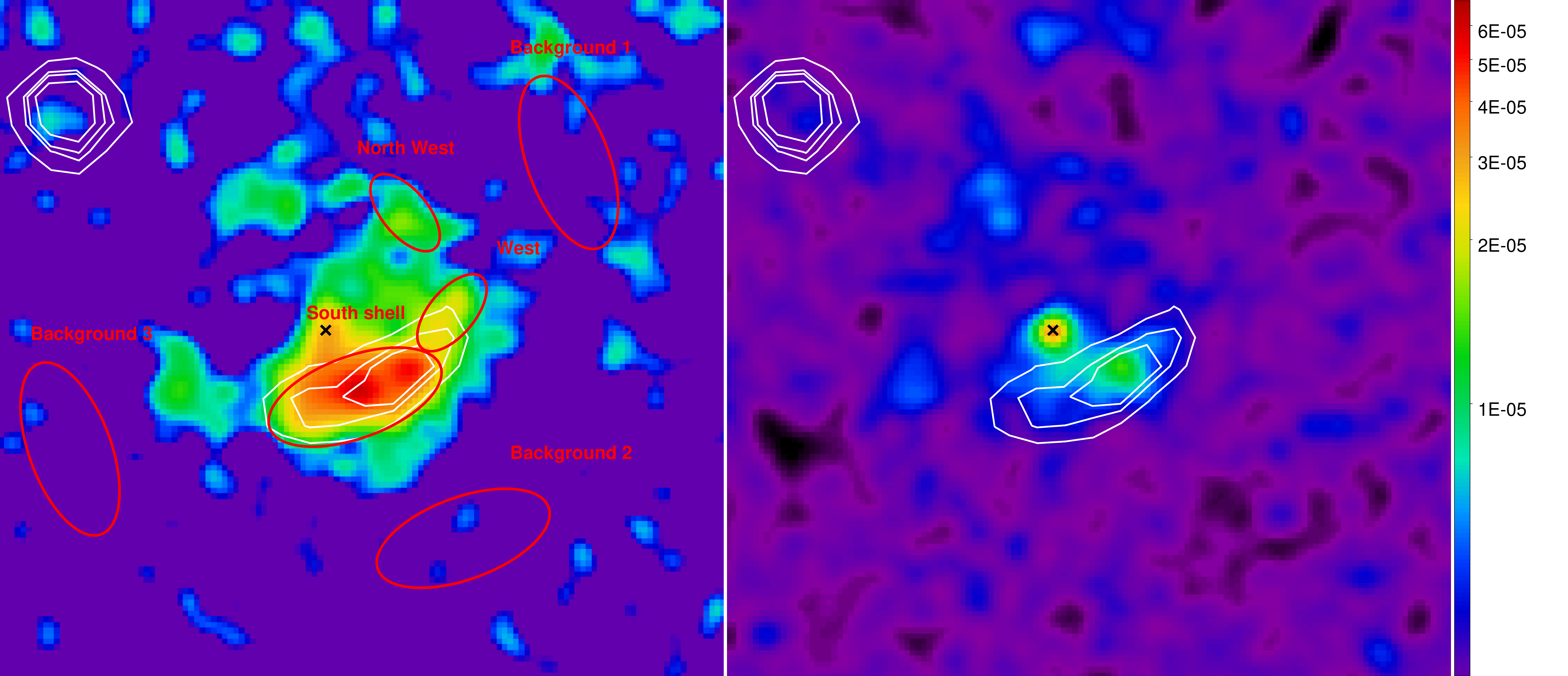}
   \caption{\xmm\ EPIC energy-resolved images of \src. The left panel shows the image in the energy range of 0.2--1\,keV revealing the prominent contribution of the SNR emission, while on the right the image in the energy range of 1--12\,keV is dominated by an unresolved source, suggested as HMXB. The optical position of the HMXB is marked with a black cross. The extraction regions for spectral analysis are shown in red and overlaid are white radio contours from the latest Australian Square Kilometre Array Pathfinder (ASKAP) survey of the LMC at 888~MHz (bandwidth is 288~MHz).} 
   \label{fig2}
\end{figure*}


 
\section{Results}
\label{Sect:results}

\subsection{Morphology of \src}
Figures~\ref{fig1} and \ref{fig2} show a faint X-ray point source at the geometrical centre of \src\,. The X-ray point source is most prominent in the hard X-rays above 1\,keV. 
The SNR displays a shell like morphology with clear structures visible in the south, west and north-west regions. 
To estimate the size of the SNR, we employed the method described by \cite{2015A&A...579A..63K}, which fits an ellipse to the outer contours of the SNR (at 3$\sigma$ above the surrounding background level in the 0.2$-$1 keV EPIC image). We derive the ellipse centre at 
R.A. = 05$^{\rm h}$13$^{\rm m}$43.0$^{\rm s}$ and Dec. = $-$67\degr24\arcmin10\farcs0 (J2000) with a positional error of 3\arcsec. The semi-major and semi-minor axes correspond to a  size of 14.5 ($\pm1.0$) $\times$ 12.3 ($\pm1.0$) pc at the distance of LMC (50 kpc) with the major axis rotated $\sim$82\degr\ East of North. 
Figure~\ref{fig1} also displays the radio contours from the latest Australian Square Kilometre Array Pathfinder \citep[ASKAP,][]{2008ExA....22..151J} survey of the LMC at 888~MHz
showing correlated emission in the radio and X-ray wavelengths from the south shell of the emission from \src. Figure~\ref{fig2}, right panel, 
further demonstrates that the brightest and hardest emission from the SNR (from the south shell) is spatially coincident with the brightest radio emission. This further testifies to its identity as an SNR.

\subsection{X-ray position of the point-source and identification of the optical/infrared counterpart}
In order to determine the X-ray position of the point source, we performed a
maximum-likelihood source detection analysis on the \xmm/EPIC images.  This was done on fifteen images created from the three EPIC cameras in five energy bands as given: $1\rightarrow(0.2-0.5$ keV),   $2\rightarrow(0.5-1.0$ keV),  3$\rightarrow(1.0-2.0 $ keV), $4\rightarrow(2.0-4.5 $ keV), $5\rightarrow(4.5-12.0$ keV) \citep{2009A&A...493..339W,2013A&A...558A...3S}. Source detection was performed simultaneously on all the images using the SAS task {\tt edetect\_chain}.  The best-determined position is R.A. = 05$^{\rm h}$13$^{\rm m}$42.59$^{\rm s}$ and Dec. = $-$67\degr24\arcmin12\farcs4 (J2000)
with a $1\sigma$ statistical uncertainty of 1.04\arcsec. The positional error is usually dominated by systematic astrometric uncertainties. Therefore, we added
a systematic error of 0.37\arcsec\ in quadrature \citep{2016A&A...590A...1R}.
The position of the point source is consistent within errors with the derived centre of the SNR.
As the source is most prominently seen in the harder X-ray bands, we repeated the source detection with the method described above, however using only the hard X-ray bands (1--2\,keV, 2--4.5\,keV and 4.5--12\,keV). The best-obtained position is R.A. = 05$^{\rm h}$13$^{\rm m}$42.65$^{\rm s}$ and Dec. = $-$67\degr24\arcmin12\farcs9 (J2000) with a slightly reduced $1\sigma$ statistical uncertainty of 0.94\arcsec. In addition, the source detection likelihood (DET$_{ML}$) increased significantly from 13.1 in the entire energy range to 47 in the hard X-ray bands. It should be noted that the point-source was not detected by performing the source detection in the energy range of 0.2--1\,keV indicating that the diffuse emission dominates in this energy range.  

To first verify whether the source is an AGN in the background, we correlated the source position with the all-sky catalogue of 1.4 million AGNs of \cite{2015ApJS..221...12S}, the Half-Million Quasars catalogue (HMQ) and MILLIQUAS catalogue of \cite{2015PASA...32...10F,2017yCat.7277....0F}. An angular separation of 
      \begin{equation}\label{eqn}
       r \leq 3.439 \times  \sqrt{\sigma_{\rm X}^2+\sigma_{\rm catalogue}^2} = 3.439\sigma
      \end{equation}
was used. For a  Rayleigh distribution this corresponds to a 3$\sigma$ completeness, but no counterpart was found. We further investigated whether the source could have an infrared counterpart satisfying the mid-infrared color selection for an AGN \citep{2012MNRAS.426.3271M}, but it has magnitudes too faint to be detected. As the X-ray and the infrared flux in AGN are expected to be correlated, as shown in Fig. 11 in \cite{2019A&A...622A..29M}, we estimated the expected magnitudes of the counterpart in the ALLWISE bands (W1, W2, W3) for the given X-ray flux of the source. An AGN counterpart, if it exists, is expected to have W1, W2 and W3 in the range of 14.6--17.5, 13.4--16.4 and 10.6--12.7 mag respectively, and should therefore be visible. The nearest ALLWISE counterpart is much brighter (see below) thereby making this possibility highly unlikely. 

The Magellanic Clouds, especially the Small Magellanic Cloud, host a large population of HMXBs owing to the relatively recent star formation history \citep{2010ApJ...716L.140A,2016A&A...586A..81H,2018MNRAS.475.3253V}. To investigate further a possible HMXB nature of the unresolved source, we searched the Magellanic Clouds Photometric Survey (MCPS) catalogue \citep{2004AJ....128.1606Z} for a counterpart using the same correlation criterion as equation~\ref{eqn}. In order to reduce the probability of a chance coincidence, we first filtered the MCPS catalogue in colour and magnitude space using the loci of the confirmed HMXBs in the LMC on the colour-magnitude and colour-colour diagram. The details of the criterion and a comprehensive list of HMXBs in the LMC will be presented in a forthcoming work. We found a unique optical counterpart within the  3$\sigma$ completeness radius with $U=12.07\pm0.03$, $V=13.4\pm0.04$, and $I=13.11\pm0.04$ mag, satisfying the colour and magnitude criteria of early-type stars in the LMC.  We also found the corresponding infrared counterpart of the source (2MASS 05134260--6724100) with  $J=12.93\pm0.03$, $V=12.84\pm0.04$, and $I=12.69\pm0.04$ mag and (ALLWISE J051342.62--672410.0) $W1=12.40\pm0.02$, $W2=12.23\pm0.02$, and $W3=9.66\pm0.06$ mag. 
The optical counterpart was identified as the south-eastern star of a close pair by \cite{1978A&AS...31..243R} and shows variability in the OGLE-III I and V bands \citep{2013AcA....63..159U}. Simbad lists the star as post-AGB star candidate (\osrc) with spectral type B2.5Ib inferred by \cite{1972A&A....18..271D} from three-colour photometry.
Since no higher-resolution spectroscopy exists from this star, we compared its broad-band spectrum based on UBVRIGJHK photometric measurements to those of wind-fed supergiant HMXBs (Vela\,X-1 and 4U\,1700-37) and Be/X-ray binaries in the Magellanic Clouds. A much better agreement of \osrc\ with the wind-fed supergiant systems is evident. We also created a colour-magnitude diagram from OGLE V-I vs. I data for the stars in the surrounding , which shows \osrc\ as one of the brightest stars at the upper end of the main sequence and redder by about 0.5 mag.
All this evidence suggests that the X-ray point source is identified with a supergiant HMXB in the LMC. Further, its positional coincidence with the geometrical centre of the SNR  asserts that it is an HMXB associated with the SNR.

\subsection{Optical light curve from OGLE}

\osrc\ was observed by the Optical Gravitational Lensing Experiment (OGLE), which started observations in 1992 \citep{1992AcA....42..253U} and continues observing till today \citep[OGLE-IV,][]{2015AcA....65....1U}. Observations are performed with the 1.3~m Warsaw telescope at Las Campanas Observatory, Chile. Images are taken in the V and I filter pass-bands and photometric magnitudes are calibrated to the standard VI system.

\begin{figure}
   \centering
   \includegraphics[width=0.47\textwidth,clip=]{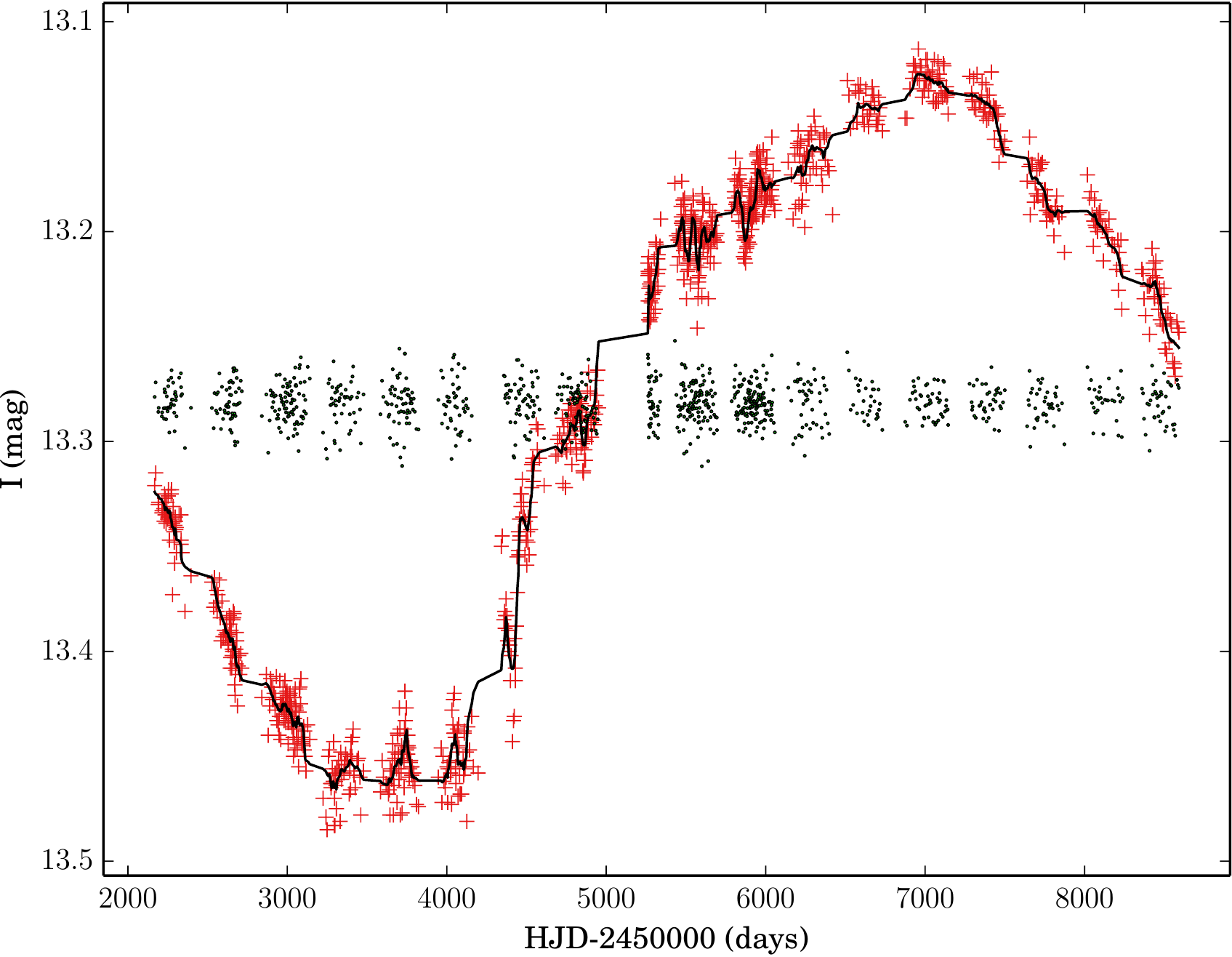}
   \caption{OGLE I-band light curve of \osrc\ (red crosses). OGLE-III measurements were obtained up to HJD 2454951 (OGLE-ID lmc108.3.23), the later data are from OGLE-IV (lmc505.07.38). The smoothed light curve is indicated by the black line. The detrended light curve after subtraction of the smoothed data is plotted with black dots, shifted by the I-band average of 13.28 mag.}
   \label{ogle-lc-detrend}
\end{figure}

\begin{figure}
   \centering
   \includegraphics[width=0.47\textwidth,clip=]{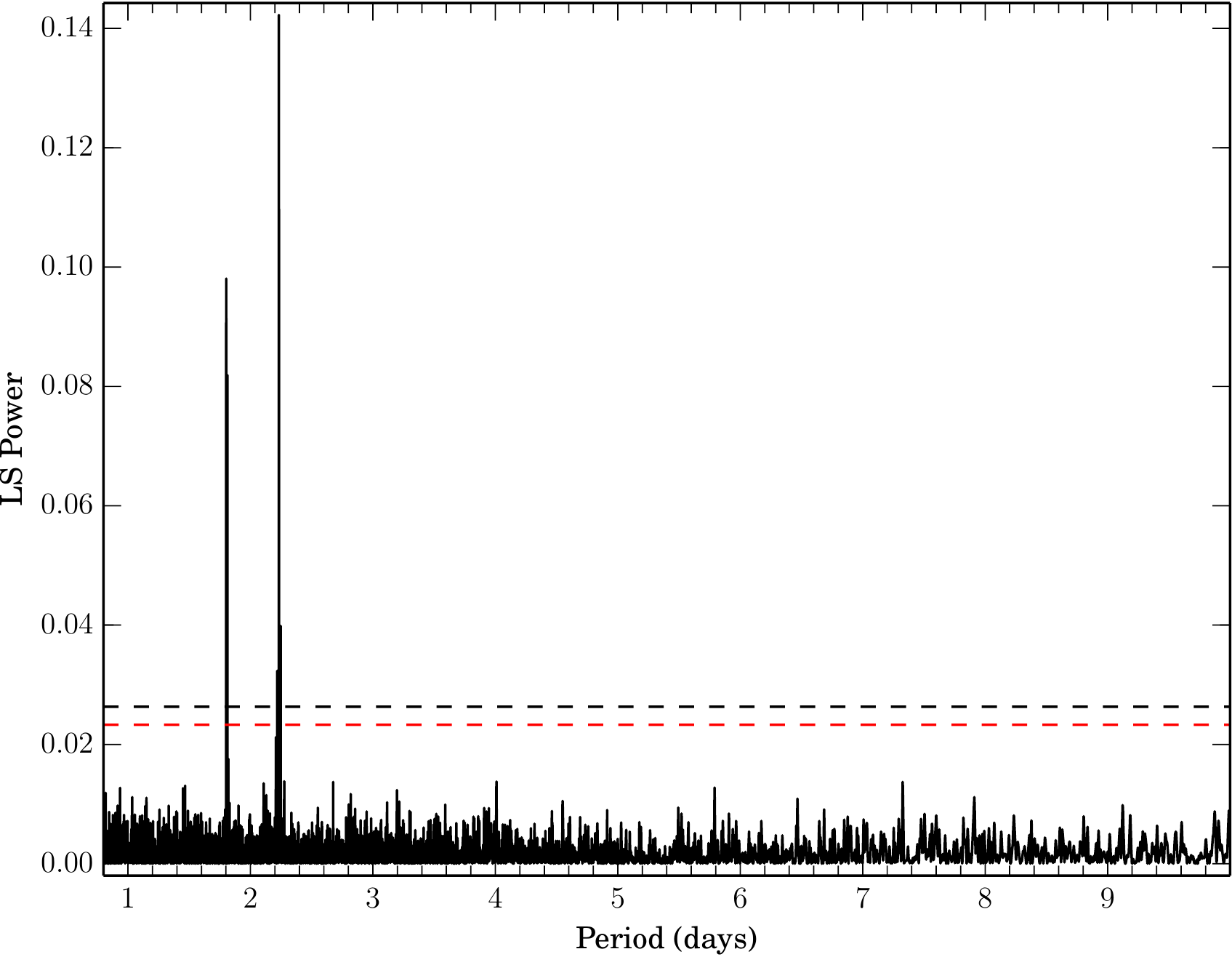}
   \caption{Lomb-Scargle periodogram of the detrended OGLE I-band light curve of \osrc. No further strong peaks are seen up to 60 d. The red and black dashed lines mark the 95\% and 99\% confidence levels, respectively.}
   \label{ogle-lc-ls}
\end{figure}

\begin{figure}
   \centering
   \includegraphics[width=0.47\textwidth,clip=]{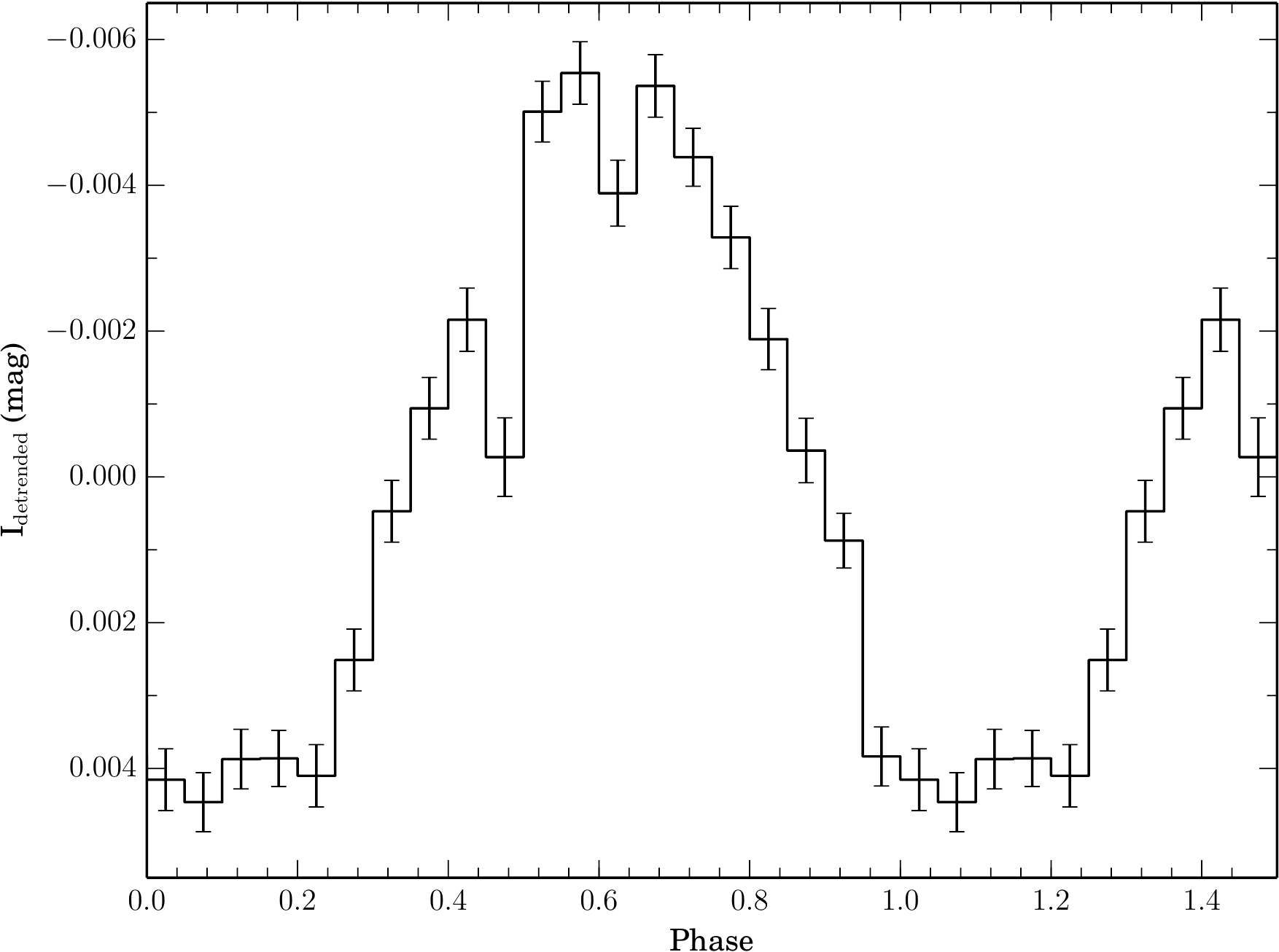}
   \caption{Detrended OGLE I-band light curve folded with a period of 2.2325\,d.}
   \label{ogle-lc-fold}
\end{figure}

Figure~\ref{ogle-lc-detrend} shows the OGLE I-band light curve of the optical counterpart obtained during observing phases III and IV. A nearly sinusoidal long-term change by more than $\pm$0.15 mag is visible over $\sim$6930 d, which is superimposed by $\sim$0.05 mag variations within the half-yearly visibility windows. To investigate the short-term variations we removed the long-term trend by subtracting a smoothed light curve \citep[derived by applying a Savitzky-Golay Filter with a window length of 13 data points;][]{1964AnaCh..36.1627S} and computing periodograms using the Lomb-Scargle algorithm \citep{1976Ap&SS..39..447L,1982ApJ...263..835S}. Two highly significant peaks are found in the periodogram at 1.8025\,d and 2.2324\,d  (Fig.~\ref{ogle-lc-ls}), which are aliases of each other with the 1-day sampling period (which also appears in the periodogram when using wider smoothing windows). Given that the stronger peak is found at 2.2325$\pm$0.0003\,d, we identify this period as the likely orbital period of the HMXB system. The detrended light curve folded with that period is presented in Fig.~\ref{ogle-lc-fold}.


\subsection{X-ray timing analysis}

\begin{figure}
   \centering
   \includegraphics[scale=0.55]{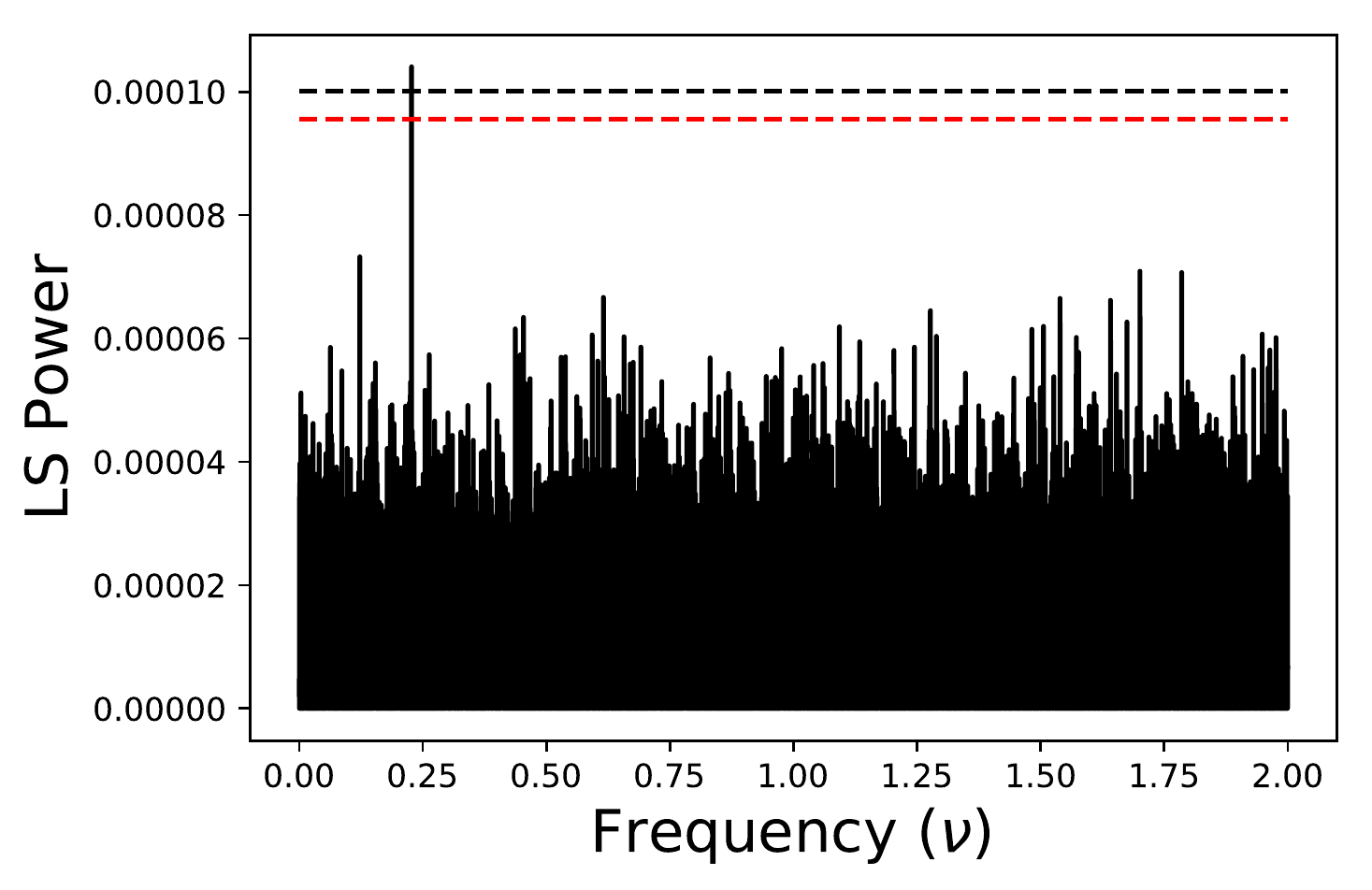}
   \caption{Lomb-Scargle periodogram of the EPIC-PN light curve in the energy band of 1--10\,keV (ObsID 0671090101). The peak indicating the spin period of a neutron star. The red and black dashed lines mark the 90\% and 95\% confidence levels, respectively.}
   \label{figtiming}
\end{figure}

\begin{figure}
\centering
\includegraphics[scale=0.28,angle=-90]{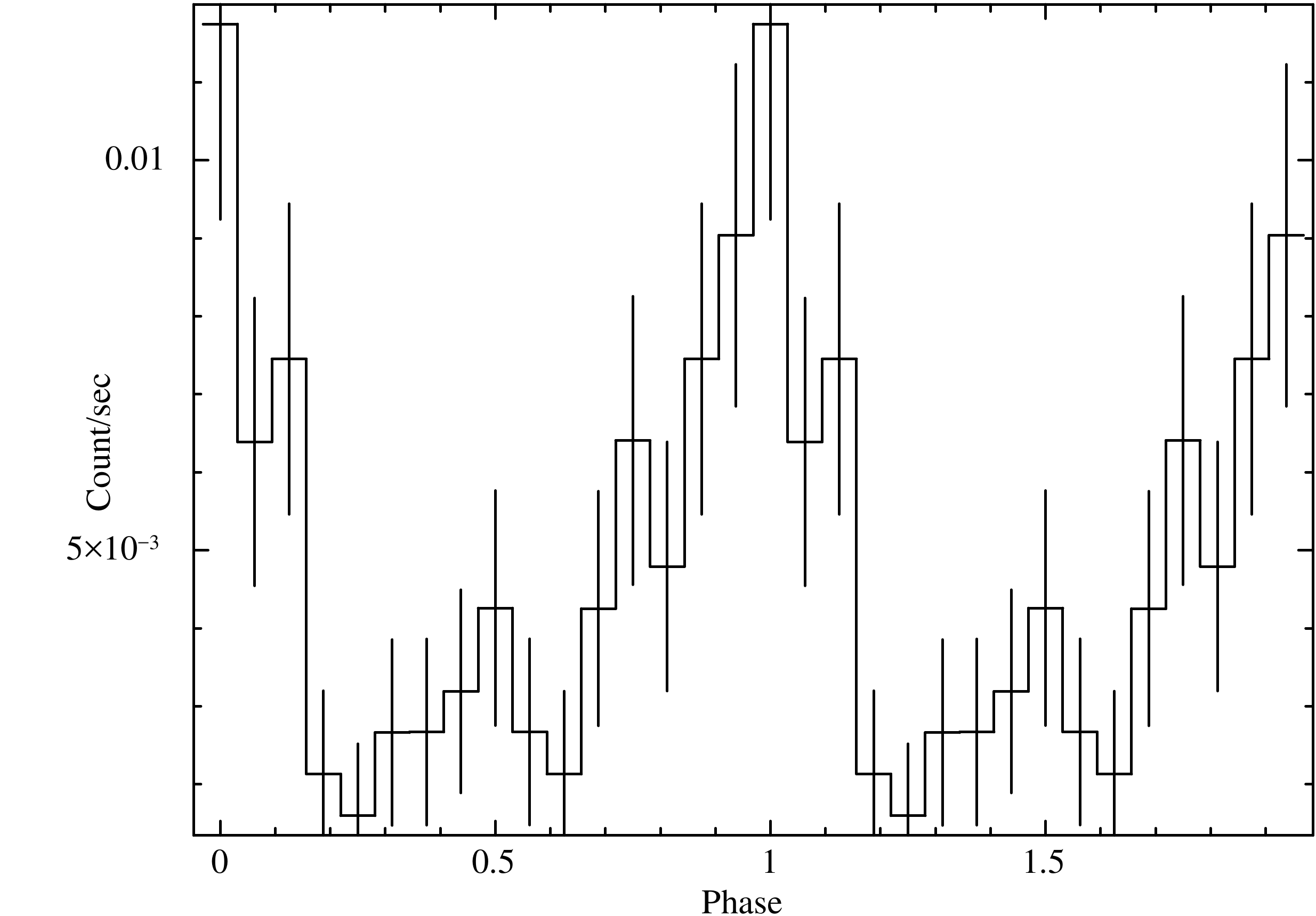}
\caption{EPIC-PN light curve folded with 4.399\,s, showing the pulse profile of the HMXB in the energy band of 1--10\,keV.}
\label{figpp}
\end{figure}

To look for a possible periodic signal in the X-ray light curve of the identified candidate HMXB, we extracted source events using a circular region with radius 10\arcsec\ centred on the best-fit position. A small extraction region was used in order to minimise the contribution of the diffuse emission from the SNR. At first, we searched for a periodic signal in the barycentre-corrected PN light curve in the energy range $>1$\,keV using a Lomb-Scargle periodogram analysis in the period range of 0.5-3000\,s \citep{1976Ap&SS..39..447L,1982ApJ...263..835S}. The highest peak in the periodogram at 4.399\,s indicates the possible spin period of the neutron star in the HMXB (Fig.~\ref{figtiming}). The confidence intervals at 90\% and 95\% are derived using the block-bootstrap method and are marked in the figure, with the peak in the periodogram corresponding to a false alarm probability of 0.04.
In order to determine the pulse period more precisely, we employed the Bayesian periodic signal detection method described by \cite{1996ApJ...473.1059G}. The spin period and its associated 1$\sigma$ error are determined to 4.39923$\pm$0.00006\,s.
As the \xmm\ EPIC data were obtained in full-frame CCD readout mode, only the PN data could be used for the purpose of timing analysis, as the MOS data do not have sufficient temporal resolution.  The EPIC-PN light curve in the range of 1--10\,keV, folded with the best-obtained period is shown in Fig.~\ref{figpp}.

\subsection{X-ray spectral analysis}
For the spectral analysis, the SAS tasks \texttt{rmfgen} and \texttt{arfgen} were used to create the redistribution matrices and ancillary files. The spectra were binned to achieve a minimum of one count per spectral bin. The spectral analysis was performed using the {\small XSPEC} fitting package, version~12.9 \citep{1996ASPC..101...17A} using the C-statistic. Errors were estimated at 90\,percent confidence intervals. For the spectral analysis of the SNR, appropriate source regions (corresponding to  the
south rim, the west rim and the north-west rim) were selected as shown in Fig.~\ref{fig2}. The instrumental/detector background can have strong spatial variations, and therefore the background spectrum was extracted by averaging over several regions across the detector plane, at similar distances to the centre of the SNR (see Fig.~\ref{fig2}). The same background was used for all the spectral analysis described in the paper. Two spectra were extracted per instrument (PN, MOS1, and MOS2) from each region (source and  background). One of them was extracted from the event list of the science observation and the other spectrum was extracted from the Filter Wheel Closed (FWC) data (which contains the information of the instrumental background component). The FWC spectra were extracted at the same detector position as in the science observation because of the strong position dependency of the instrumental background.
The corresponding spectra from the FWC data were subtracted from the science spectra of the source and background regions to subtract the quiescent particle background component.  

For the spectral modelling of the SNR, the source spectra for each of the extracted regions of \src\  and the background spectrum (as shown in Fig.~\ref{fig2}) from 
all the available instruments (PN, MOS1, and MOS2) were fitted to constrain the source component as well as the astrophysical background (AXB) component. 
To model the AXB we followed the same procedure as given in \cite{KS10}. The first AXB component is an unabsorbed thermal component for the Local Hot Bubble
(LHB, $kT\sim$0.1\,keV) while the second component is an absorbed thermal component for the
Galactic halo emission ($kT\sim$0.25\,keV). Both of these components were fitted with the 
{\small XSPEC} model APEC with elemental abundances frozen to solar values. Lastly, the third component is an absorbed 
power law ($\Gamma=1.46$) for the non-thermal unresolved extra-galactic X-ray background. The normalisations of all three were set free. It was assumed
that the temperature of the thermal components and the surface brightness of all the AXB
components do not vary significantly between the source and background regions. Thus, the appropriate temperature and normalisation parameters were linked between the two.
For the spectral analysis of the identified HMXB, the same region was used for spectral extraction as for the timing analysis. The foreground column density
$N_{H Gal}$ at the location of \src\ was taken (and fixed) from the HI maps of \cite{1990ARA&A..28..215D}.  No additional absorption component denoting the absorption in the LMC or a local absorption could be added to the spectrum given its statistical quality, and therefore the inability to constrain several low energy components in the spectrum. The solar element abundances were taken from \cite{2000ApJ...542..914W}.
\begin{table*}[t]
  \begin{center}
    \caption{X-ray spectral parameters of \src\ for the best-fit SEDOV model}
    \label{SpectralFitParametersTable}
    \begin{tabular}{lcccc}
    \hline
    Region & South shell & West & North West \\
    \hline
    Net count rate (c/s) & 0.01 (M1+M2) & 0.002 (M1+M2) & 0.002 (PN+M1+M2) \\
    Shock/electron temperature (keV) & 2.2$^{+1.2}_{-1.1}$ & 2.2 (frozen) & 2.2 (frozen)   \\
    Ionization timescale $\tau$ ($10^{10}$ s cm$^{-3}$) & $5.9^{+5.5}_{-2.2}$ & 5.9 (frozen) & 5.9 (frozen) \\
    Normalization (cm$^{-5}$) & $3.7^{+1.5}_{-0.8}$$\times$10$^{-6}$ &  $0.50^{+0.30}_{-0.20}$$\times$10$^{-6}$ & $0.20^{+0.2}_{-0.10}$$\times$10$^{-6}$\\
    \hline
    C-Statistic        & 270.71 & 121.16 & 233.71 \\
    Degrees of Freedom & 383    & 192    & 306    \\
    \hline
    \multicolumn{5}{l}{{\bf Notes:} All quoted error bounds correspond to the 90\% confidence levels. In the case of the APEC} \\ 
    \multicolumn{5}{l}{model, the normalization is defined as (10$^{-14}$/4$\pi$$d$$^2$)$\int$$n$$_{e}$$n$$_{p}$$d$$V$, where $d$ is the distance to the} \\ 
    \multicolumn{5}{l}{SNR (in units of cm), $n$$_{\rm{e}}$ and $n$$_{\rm{p}}$ are the number densities of electrons and protons respectively} \\
    \multicolumn{5}{l}{(in units of cm$^{-3}$), and $\int$$d$$V$ = $V$ is the integral over the entire volume (in units of cm$^3$).}
    \end{tabular}
  \end{center}
\end{table*}

\subsubsection{SNR}
We fitted the X-ray spectra of the SNR with different thermal emission models, typical for shock-heated plasma in SNRs. We further checked that a simple power-law model did not provide an adequate description of the spectrum, which would be indicative of non-thermal X-ray emission from the SNR.
Due to the relatively broad point spread function of the \xmm\ telescopes, we estimated that $\sim$17\%, 10\% and 11\% of the source counts from the point source will fall into the south, west and north-west regions of the SNR shown in Fig.~\ref{fig2}. We therefore included an appropriately scaled model of the point source HMXB into the spectral model of the SNR.  

\begin{figure}
\centering
\includegraphics[scale=0.3,angle=-90]{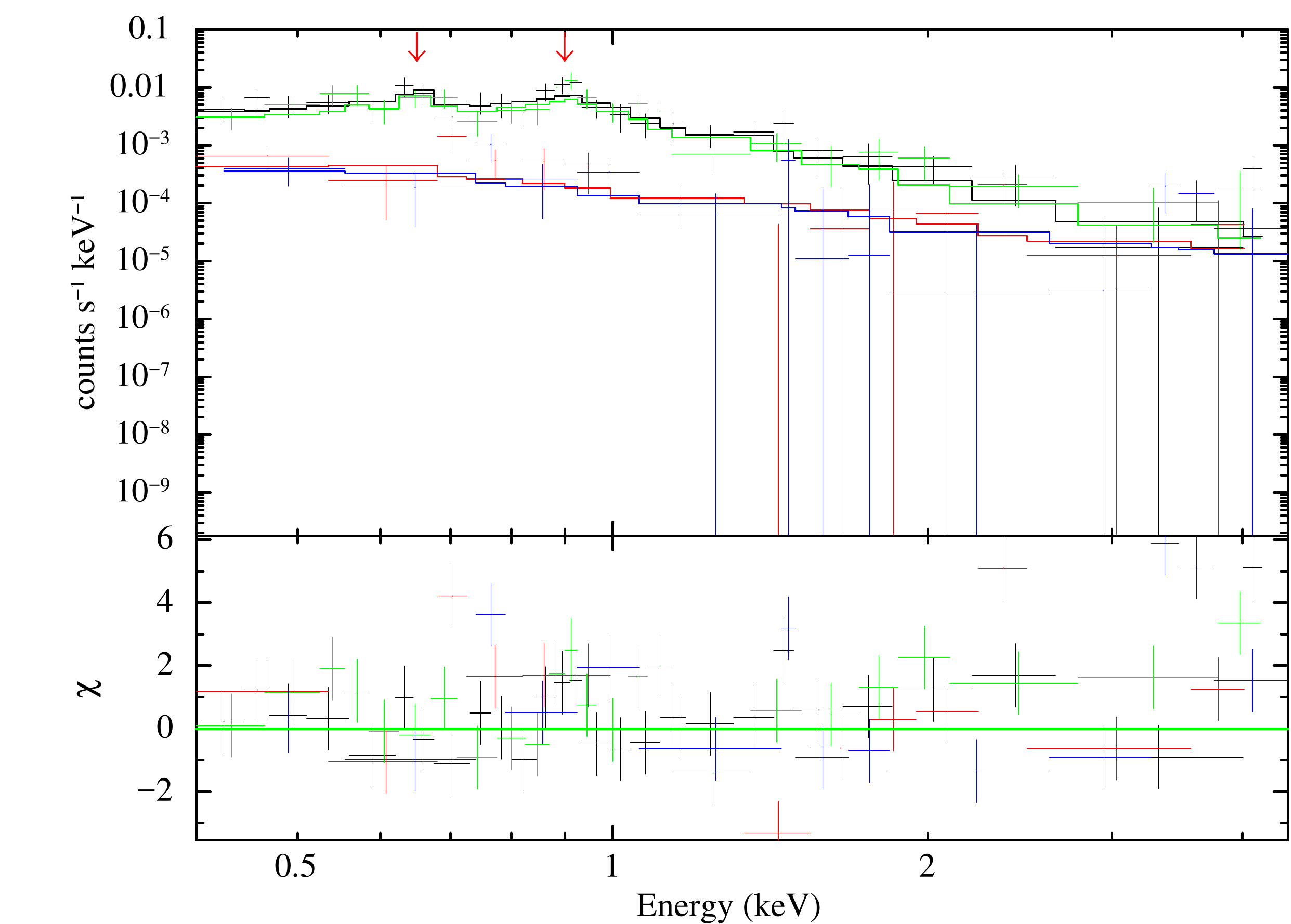}
\includegraphics[angle=-90,scale=0.3]{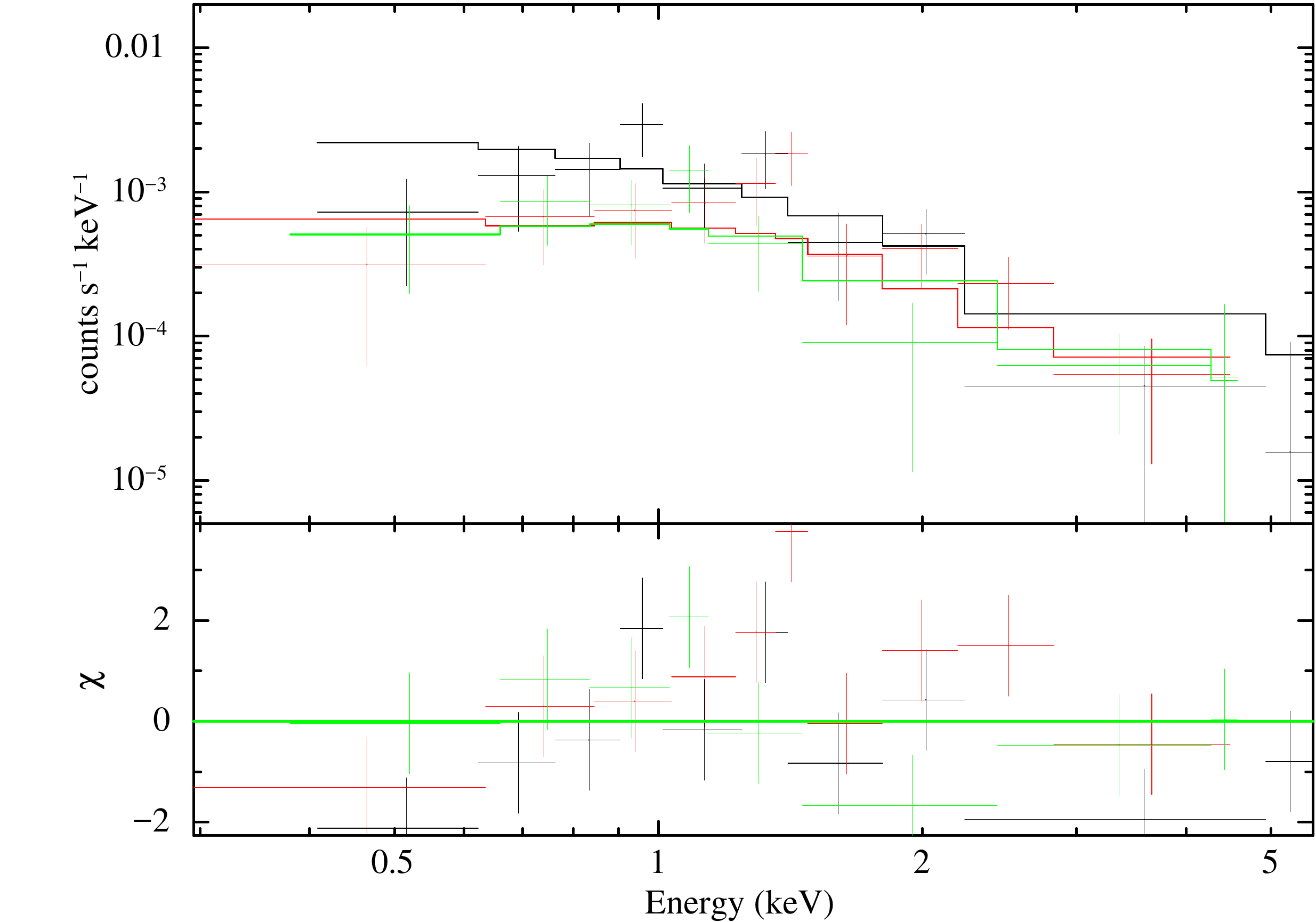}
\caption{Top: EPIC-MOS spectra of the south shell of the \src\ with best-fit VSEDOV model. The black and green data points represent the MOS1 and MOS2 spectra from the source region, the red and blue data points denote the AXB. The corresponding best-fit models are shown in the same colours. The presence of O and Ne lines are shown with arrows.
Bottom: EPIC spectra of the HMXB candidate. Black, red and green denote PN, MOS1 and MOS2 data points and model (histogram), respectively. The spectra have been rebinned for visual clarity.}
\label{figsnrsouth}
\end{figure}

An attempt to fit the spectra with an equilibrium ionization model like
APEC \citep{2001ApJ...556L..91S} provided a statistically unacceptable fit. On the other hand, non-equilibrium shock models such as PSHOCK, SEDOV \citep{2001ApJ...548..820B} provided good fits to the data as is expected from young/less evolved SNRs. Both the PSHOCK, SEDOV models have the same number of degrees of freedom and the latter one provides a marginally better fit ($\Delta\chi^{2}=7$). The best-fit spectra along with their best-fit models and residuals are shown in Fig.~\ref{figsnrsouth} and the best-fit parameters are listed in Table~\ref{SpectralFitParametersTable}. As the PN data of the south and west of the SNR fell in a CCD gap, only the MOS data were used. The best-fit temperature obtained from the highest signal-to-noise spectrum (west rim) of  2.2$^{+1.2}_{-1.1}$\,keV and ionization age of $5.9^{+5.5}_{-2.2}$ $\times10^{10}$ s cm$^{-3}$ indicates a young/less evolved SNR \citep{Vink2016}. Due to the poor statistical quality of the data, the abundances could not be set free in the spectral model, and were therefore fixed to the LMC ISM abundance \citep{1992ApJ...384..508R}. 
However, the SNR spectra in Fig.~\ref{figsnrsouth} clearly show the presence of O and Ne lines,  indicating a core-collapse supernova explosion, if interpreted as of supernova ejecta origin. For the west and the north-west rim regions of the SNR, the temperature and ionization timescales could not be constrained due to limited counts in the spectra and were therefore frozen to the best-fit value from the south shell of the SNR. Only the normalisations were set free. The total observed luminosity of the system (SNR+HMXB) is $\sim 1.7\times10^{34}$\, \ergs. A comparison with the X-ray luminosity function of the LMC SNRs  show that \src\ lies in the faint end of the distribution with the luminosities of the LMC SNR population spanning from $\sim7\times10^{33}$\, \ergs -- $3.2\times10^{37}$\, \ergs \citep{2016A&A...585A.162M,2019arXiv190811234M}.

As a radio source, \src\ is characterized by a spectral index  $\alpha=-0.68\pm0.04$, a flux density of S$_{\rm 888\,MHz}=45\pm3$~mJy, a surface brightness  $\Sigma=1.0\times10^{-21}$\,W\,m$^{-2}$\,Hz$^{-1}$\,sr$^{-1}$, and a total radio luminosity (10~MHz -- 100~GHz) of 1.58$\times10^{28}$\,W (Filipovi{\'c} et al., in preparation). This would position \src\ among the least luminous LMC SNRs \citep{Bozzetto2017}. 

\subsubsection{HMXB}
The X-ray spectrum from the point source contains a significant contribution from the underlying extended SNR as can be seen from Fig.~\ref{fig2}. In order to model this contribution, we included a component for the SNR emission in the spectral fit with the normalization set free. The spectrum of the HMXB was modelled with a power-law.  
The spectra along with their best-fit model and residuals are shown in Fig.~\ref{figsnrsouth} and the best-fit parameters are listed in Table.~\ref{tabspec}. The source is faint with an absorption-corrected luminosity of $7.3^{+2.6}_{-1.5}$~$\times~$10$^{33}$ erg~s$^{-1}$ (0.2-12\,keV).

\begin{table}
\caption{X-ray spectral parameters of the identified HMXB in \src\ for a power-law model.}
\begin{tabular}{lc}
   \hline
   Parameter & Value \\
   \hline
   Net count rate (c/s) & 0.004 (PN+M1+M2) \\
   $\Gamma$ & 1.63$\pm0.29$ \\
   Absorption corrected X-ray luminosity$^{a}$ (erg~s$^{-1}$) & $7.3^{+2.6}_{-1.5}$~$\times~$10$^{33}$ \\
   \hline
   C-Statistic        & 124.6  \\
   Degrees of Freedom & 126    \\
   \hline
\end{tabular}

$^{a}$Assuming a distance of 50\,kpc in the energy band of 0.2--12\,keV\,. 
The absorption column density was fixed to the Galactic foreground value of N$_{\rm Hgal}$=~6$\times$10$^{20}$\,cm$^{-2}$. 
Errors are quoted at 90\% confidence.
\label{tabspec}
\end{table}

\section{Discussions and conclusions}
\label{Sect:discussion}
We report the discovery of a new HMXB system and its associated supernova remnant \src. 
The HMXB is X-ray faint with a probable NS spin period of 4.4\,s.
As optical counterpart we identify \osrc, a supergiant with spectral type B2.5Ib which needs further confirmation with high-resolution optical spectroscopic observations. 
From the I-band light curve extending over more than 17.5\,years, we find significant periodic variations with 2.2324$\pm$0.0003\,d, which we interpret as the orbital period of the HMXB system. The temperature and ionization timescales of the SNR as obtained from the spectral analysis indicate a young/less evolved SNR with an age of $<$6\,kyr. 
There are three other confirmed HMXBs detected in the Magellanic Clouds known to have SNR associations. The two systems located in the Small Magellanic Cloud, a.k.a  SXP\,1062 and SXP\,1323 are Be X-ray binaries (BeXRBs) which are a different class of HMXBs. BeXRBs have wider, more eccentric orbits, and the compact object accretes from the circumstellar disc of the Be star. These systems follow a different evolutionary path in comparison to supergiant HMXBs \citep[see e.g.][]{2011ASPC..447...29C}. The age of SXP\,1062 is estimated to be $\sim$16\,kyr \citep{2012A&A...537L...1H}, and SXP\,1323 is estimated to $\sim$40\,kyr \citep{2019MNRAS.485L...6G} indicating older, middle aged SNRs. Furthermore, the NS in these systems have very long spin periods, in contrast to the HMXB in \src. 
DEM\,L241 is a HMXB-SNR system located in the LMC 
with a probable NS as the compact object \citep{2019MNRAS.484.4347V} and an O-type companion star. The age of the SNR is estimated to be $>50-70$ kyr, i.e. also much larger than \src\, \citep{2012ApJ...759..123S}.

\subsection{Nature of the compact object} 
The spatial association of the X-ray point source with a massive early-type star (probable supergiant), its hard X-ray spectrum and the detection of pulsations all point to an HMXB with an NS as compact object. 

\subsection{Nature of the SNR and its properties}
The detection of the SNR at energies $>1$\,keV, and the ionization timescale both indicate shock heated plasma from a young/less evolved SNR that is in a non-equilibrium ionization state. The elemental abundances of the plasma could not be measured using the currently existing observations and a detailed investigation of the plasma properties solicits a deeper 
observation with \xmm\ or \cxo.
The radius of the SNR is measured to be 14.5 pc (considering the radius as the semi-major axis of the best-fit ellipse to the SNR contours). In order to estimate the age of the SNR we used the equation from \cite{2005ChJAA...5..165X} using the Sedov relation \citep{1959sdmm.book.....S}. With the measured SNR radius and shock temperature we derive a dynamical age of $3700^{1900}_{-900}$ yr ($<$5600 yr). Following the method of \cite{2001ApJ...548..820B}, \cite{2011A&A...530A.132O}
and the normalization obtained from the VSEDOV model, we obtained a dynamical age of $3800^{+1900}_{-900}$ yr ($<$5700 yr). The dynamic age derived from two independent estimates are consistent and indicate a young/less evolved SNR with dynamical age $<$6\,kyr. We note that the assumption of a Sedov phase is a conservative one, as the early expansion of the SNR is un-decelerated. This provides a lower limit of $\sim 1$~kyr if the expansion has been at a constant shock speed measured by the X-ray temperature (13\,500~km~s$^{-1}$). Further, following the method of \cite{2001ApJ...548..820B}, with the obtained normalization of the Sedov model, and assuming baryon number per hydrogen atom ($r_m\approx1.4$, assuming a helium/hydrogen ratio of 0.1), we derived the ambient electron density ($n_e$) to be $\sim 0.02$~cm$^{-3}$. This is consistent with the typical values in the LMC \citep{2019arXiv190811234M}.

\subsection{The youngest known HMXB caught at the onset of accretion?}
The youngest known XRB inside an SNR is Circinus\,X-1 with an estimated age of $<$4600\,yr. This source has been commonly proposed to be a low mass X-ray binary as it exhibits type I X-ray bursts \citep{1986MNRAS.219..871T,2010ApJ...719L..84L}. Type I X-ray bursts are produced by thermonuclear explosions on the surface of an accreting NS in a low-mass X-ray binary system (see e.g. \cite{2006csxs.book..113S}). However, recent investigations indicate that a high mass companion for Circinus\,X-1 cannot be ruled out \citep{10.1093/mnras/stv2669,2019HEAD...1711233S}.
The point-source inside \src\ seems to be one of the youngest HMXB known to date. 

The typical age of NSs (time since the supernova explosion) identified as HMXBs is expected to exceed $\sim$10$^{5}$ years \citep{1991PhR...203....1B}. However, with the discovery of SXP\,1062 \citep{2012A&A...537L...1H}, and SXP\,1323 \citep{2019MNRAS.485L...6G}, it is now known that the age of NSs in HMXBs can be significantly smaller. An NS formed in an HMXB spins down from its initial short spin period up to a critical value, only after which the accretion can commence onto its surface. A nascent neutron star is first believed to be in the ejector phase where the relativistic winds prevent the matter to penetrate the Bondi radius, and hence accrete onto the NS. As the NS spins down by magnetic dipole radiation, it next enters the propeller phase, where it now spins down due to the interaction between the magnetosphere and infalling matter from the evolving companion star. The system can switch on as an accreting pulsar only after the spin period has reached a critical value, where the magnetospheric radius equals the co-rotation radius of the NS. The magnetospheric radius is a function of the mass accretion rate and magnetic field strength, and the co-rotation period mainly depends on the spin period. Hence assuming the unabsorbed luminosity (0.2--12\,keV) estimated from the HMXB-NS spectral fit as the limiting luminosity before the onset of the propeller regime, and the observed spin period of 4.4\,s as the critical value, the dipole magnetic field strength of the NS can be estimated to $\sim$3$\times10^{11}$~G \citep{2002ApJ...580..389C}. Here we used $R_{M}=0.5R_{A}$ (magnetic radius $R_{M}$ and Alfv{\'e}n radius $R_{A}$) as is widely accepted for most disc-magnetosphere models and has been recently supported by a broad set of observations by \cite{2018A&A...610A..46C}. Assuming a correction factor of 2 for the bolometric luminosity, the upper-limit of the surface magnetic field strength is $<$5$\times10^{11}$~G. The estimated field strength is low for a newly born NS which is expected to be strongly magnetized. 

The results obtained above can imply several interesting possibilities. Firstly, the NS can have a significant multi-polar component of the magnetic field which is not sampled when estimating the magnetic field strength from the above method. Alternatively, the NS can be born with a low magnetic field strength and as a slow-rotator, due to which the magnetic fields cannot be sufficiently amplified \citep{1993ApJ...408..194T}. However, later studies have showed that the magnetic field can be amplified even in the absence of fast rotation \citep{2012ApJ...751...26E}.
The third, and the most interesting possibility is that the NS is born as strongly magnetized, but its field is buried deep beneath the surface \citep{1990Natur.347..741R}, possibly by an episode of post-supernova hyper-critical accretion \citep{1989ApJ...346..847C,1999A&A...345..847G,2012ApJ...748..148S}.
These fields then diffuse out on a timescale of $10^{3}-10^{4}$ years, and a much smaller value of surface magnetic field strength is therefore observed \citep{2011MNRAS.414.2567H}. 
This phenomenon is typically observed in central compact objects (CCOs) - isolated NSs found inside SNRs - \cite{2010ApJ...724.1316G,2010ApJ...709..436H,2011ApJ...733L..28H} and also in some low-field magnetars \citep{Turolla:1532705}. 
The discovery of a very young HMXB of age $<$10$^{4}$ years can provide us the unique opportunity to observe the evolution of the observable magnetic field for the first time in XRBs. The observed X-ray luminosity of this system is also similar to that seen in CCOs, however with a different spectrum due to the effect of accretion from the companion star. In contrast, CCOs have a thermal spectrum originating from the surface of the NS \citep[see e.g. ][]{2017JPhCS.932a2006D}.

\src\ is also a potential gamma-ray binary candidate. Gamma-ray binaries, which are characterised by non-thermal emission peaking above 1\,MeV, are compact objects in orbit with a massive companion. These systems are usually thought to be a short-lived phase in the evolution of HMXBs, preceding the phase when the accretion commences on to the compact object \citep{2006csxs.book..623T,2017A&A...608A..59D}. The only gamma-ray binary in the LMC known to date is DEM\,L241 \citep[LMC P3,][]{2016ApJ...829..105C} which is also a XRB-SNR system as discussed earlier. The young HMXB in \src\, is in tune with the above scenario, and deserves future follow-up in gamma-rays.

\section*{Acknowledgements}
This work is based on observations obtained with XMM-Newton, an ESA science mission with instruments and contributions directly funded by ESA Member States and NASA.
The Australian SKA Pathfinder (ASKAP) is part of the Australian Telescope which is funded by the Commonwealth of Australia for operation as National Facility managed by CSIRO. We used the \textsc{karma} and \textsc{miriad} software packages developed by the ATNF. Operation of ASKAP is funded by the Australian Government with support from the National Collaborative Research Infrastructure Strategy. ASKAP uses the resources of the Pawsey Supercomputing Centre. Establishment of ASKAP, the Murchison Radio--astronomy Observatory and the Pawsey Supercomputing Centre are initiatives of the Australian Government, with support from the Government of Western Australia and the Science and Industry Endowment Fund. We acknowledge the Wajarri Yamatji people as the traditional owners of the Observatory site.
The OGLE project has received funding from the National Science Centre, Poland, grant MAESTRO 2014/14/A/ST9/00121 to AU. M. S. acknowledges support by the DFG trough the grant SA 2131/12-1



\bibliographystyle{mnras}
\bibliography{example} 






\bsp	
\label{lastpage}

\end{document}